\documentclass[
10pt,
twocolumn,          
prl,                         
superscriptaddress
]{revtex4-2}

\usepackage{amsmath,amssymb,bm}
\usepackage{braket}
\usepackage{graphicx}
\usepackage[version=4]{mhchem}
\usepackage{mathrsfs}
\usepackage{xcolor}
\usepackage{soul}

\usepackage[
colorlinks=true,
linkcolor=blue,
citecolor=blue,
urlcolor=blue
]{hyperref}

\begin{document}

\title{Dynamic chirality in photonic time crystals}

    \author{Jianming Mai}
	\affiliation{Department of Physics, City University of Hong Kong, Kowloon, Hong Kong, China}
	
    \author{Chenwen Yang}
	\affiliation{Department of Physics, City University of Hong Kong, Kowloon, Hong Kong, China}
	
	\author{Shubo Wang}
	\email{Email: shubwang@cityu.edu.hk}
	\affiliation{Department of Physics, City University of Hong Kong, Kowloon, Hong Kong, China}




\begin{abstract} 
Temporal modulation offers a fundamentally distinct degree of freedom for active wave control beyond static spatial structuring. Photonic time crystals (PTCs), based on periodic modulation of electromagnetic parameters in time, have expanded photonic band engineering from space to time by enabling controlled energy exchange between light and the modulation. Yet, the use of PTCs to synthesize rotational dynamics and thereby control chirality and circular dichroism (CD) remains largely unexplored. Here, we propose a spatiotemporal PTC whose central cylindrical element is driven by an azimuthally traveling-wave permittivity modulation. Although the structure is geometrically static, its dielectric profile evolves as an effectively rotating pattern in time. This synthetic rotation lifts a static modal degeneracy and produces two nondegenerate counter-rotating states with opposite orbital angular momenta. These chiral modes selectively couple to left- and right-circularly polarized light, giving rise to tunable CD. In addition, the spatiotemporal modulation induces orbital angular-momentum conversion between the Floquet replica bands. Our work reveals the microscopic origin of dynamic chiral response and establishes a strategy for reconfigurable chiral photonics without mechanical motion.

\end{abstract}
\maketitle

Conventional artificial electromagnetic materials, such as metamaterials and photonic crystals, manipulate light primarily through spatial structuring. Temporal modulation departs from this static paradigm by rendering electromagnetic parameters explicitly time-dependent, opening an additional temporal design dimension for controlling wave propagation, scattering, and resonant responses \cite{lustig2018topological,park2022revealing,galiffi2022photonics,guo2026plasmonic,lustig2023photonic,kim2023unidirectional, wang2025expanding}. At temporal interfaces, where the medium is abruptly changed in time but remains uniform in space, waves can exchange energy with the modulation while conserving momentum\cite{lyubarov2022amplified,galiffi2023broadband}, leading to temporal refraction and reflection\cite{bohn2021spatiotemporal,xiao2014reflection,zhou2020broadband,moussa2023observation}, frequency conversion\cite{rizza2024harnessing}, negative refraction\cite{lasri2023temporal}, polarization conversion\cite{yin2022temporal}, and spin control\cite{mostafa2023spin,rizza2023spin}.

Extending this concept to periodic temporal modulation gives rise to photonic time crystals (PTCs), which are naturally described within a Floquet framework and exhibit quasifrequency band structures\cite{sharabi2022spatiotemporal,park2025spontaneous,yang2025topologically,zhang2025floquet,rudner2020band,garg2026photonic}. More generally, simultaneous periodic modulations in space and time enable spatiotemporal PTCs, offering richer wave dynamics than purely spatial or temporal counterparts, such as nontrivial topology\cite{serra2023rotating,gao2025topological,segal2025two,jin2025towards}, wave amplification and lasing\cite{huang2026microwave,yang2023cascaded}, and nonreciprocity\cite{galiffi2019broadband,chamanara2016optical,park2021spatiotemporal,huidobro2021homogenization,huidobro2019fresnel}. Beyond photonics, related time-modulation strategies have been implemented in acoustic\cite{tong2025observation,zhu2023effective}, water-wave\cite{bacot2016time}, and elastic-wave systems\cite{trainiti2019time}, suggesting a broader route to tailoring conservation laws, band structures, and scattering processes across wave platforms.

Among the functionalities enabled by spatiotemporal modulation, the control of optical chirality has remained largely unexplored. Chiral light–matter interactions underpin a variety of intriguing phenomena and functionalities, including circular dichroism (CD) \cite{plum2009metamaterials,tang2010optical}, chirality sorting \cite{wang2014lateral,shi2023advances}, geometric phases \cite{bomzon2002space,fu2024near,cheng2025riemann}, and spin-dependent light routing \cite{bliokh2015spin,wang2019arbitrary,cheng2023directional}. Such interactions are conventionally engineered through structural or material asymmetry. An alternative and intuitive route is physical rotation, which breaks the equivalence between counter-rotating states carrying opposite angular momenta. Mechanically rotating photonic structures have been shown to exhibit distinct responses to left- and right-circularly polarized light (LCP and RCP, respectively) \cite{pan2019circular,yang2025nonreciprocal,li2025optical,li2026strong}. However, physical rotation introduces mechanical complexity and is difficult to implement at high speeds or in integrated photonic platforms. This naturally raises the question of whether rotation can instead be synthesized through spatiotemporal modulation in a geometrically stationary structure, thereby enabling dynamically reconfigurable chiral light–matter interactions without actual mechanical motion.

Here we propose a spatiotemporal PTC whose central cylindrical element features an azimuthally traveling-wave permittivity modulation. Although the structure is geometrically static, its dielectric profile evolves as an effectively rotating pattern in time. This synthetic rotation lifts the degeneracy of two orthogonal modes in the static limit, giving rise to a pair of counter-rotating states carrying opposite orbital angular momenta. These chiral modes selectively couple to LCP and RCP light, thereby producing tunable CD. The spatiotemporal modulation further enables orbital-angular-momentum conversion among the Floquet replica bands. Our work elucidates the physical mechanism underlying dynamic chiral response and establishes a general strategy for realizing reconfigurable chiral photonics without mechanical motion.

\begin{figure}[tb]
\centering\includegraphics[width=\linewidth]{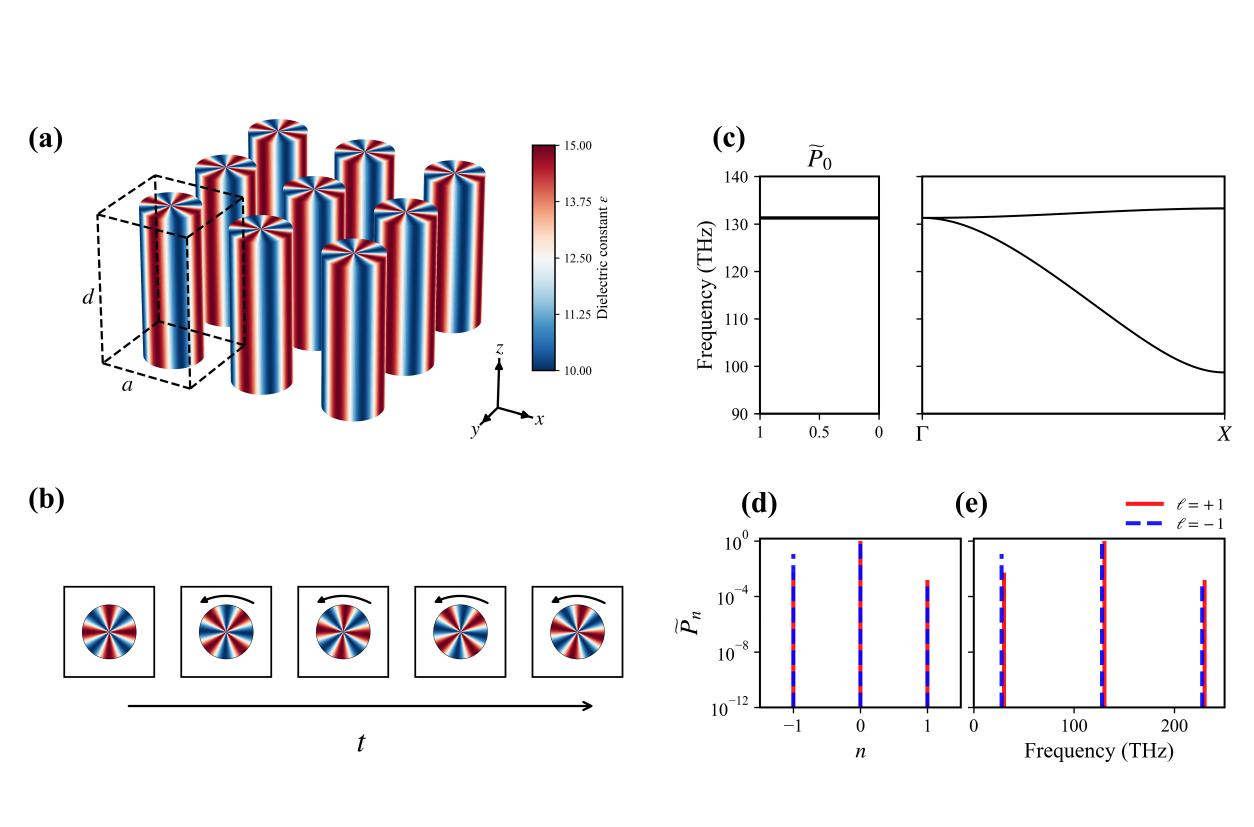}
\caption{The spatiotemporal PTC. 
(a) Schematic of the modulated crystal array and its dielectric-permittivity distribution. The periodic lattice forms a PTC whose material response is engineered through a prescribed spatiotemporal modulation. The parameters are set as \(a = 900\,\mathrm{nm}\), \(R = 0.3a\), \(d = 2.5\,\mu\mathrm{m}\), \(\bar{\varepsilon} = 12.5\), \(\mu = 1\), \(\delta\varepsilon = 0.2\), \(\varepsilon'' = 0.002\), and \(m = 4\). 
(b) Top view of a single unit cell, illustrating the time-dependent evolution of the dielectric-permittivity profile induced by the spatiotemporal modulation. The arrows denote the direction of modulation. 
(c) Static spectrum calculated using the simplified two-dimensional model. The left panel shows the normalized zeroth-order Floquet-channel population \(\widetilde{P}_0\) at the \(\Gamma\) point, identifying a doubly degenerate eigenfrequency in the static system. The right panel shows the corresponding band dispersion along the \(\Gamma\)-X direction, where the two branches are degenerate at \(\Gamma\) and split away from \(\Gamma\). 
(d) Normalized Floquet-channel populations \(\widetilde{P}_n\) of the two split chiral branches for the \(n=-1,0,+1\) Floquet orders with \(f_m=100\,\mathrm{THz}\). 
(e) Frequency-resolved sideband spectra obtained by plotting the \(\widetilde{P}_{-1}\), \(\widetilde{P}_{0}\), and \(\widetilde{P}_{+1}\) at their corresponding physical frequencies \(f_n=f_0+n f_m\). The labels \(\ell=+1\) and \(\ell=-1\) denote the two counter-rotating chiral branches originating from the static degenerate doublet, anticipating their identification as orbital-angular-momentum states below.} 
\label{fig:1}
\end{figure}

\textit{Floquet-Bloch theory}---We consider a PTC consisting of a square lattice of dielectric cylinders with lattice constant \(a\). As shown in Fig.~\ref{fig:1}(a), each unit cell contains a cylinder of radius \(R\) and height \(d\), centered in the unit cell with its axis oriented along the \(z\) direction. The cylinders are embedded in air and are assumed to be nonmagnetic. The cylinder material is taken to be weakly lossy, and the real part of the permittivity is modulated azimuthally in the transverse plane as: 
\begin{equation}
\varepsilon(\theta,t)=\bar{\varepsilon}\left[1+\delta\varepsilon\cos(m\theta-\Omega t)\right]+i\varepsilon''
\label{eq1}
\end{equation}
where \(\bar{\varepsilon}\) denotes the mean real permittivity of the cylinder, \(\delta\varepsilon\) is the relative modulation depth, \(\varepsilon''\) is the material loss, \(\theta\) is the azimuthal coordinate in the transverse plane and \(t\) is the time. The integer \(m\) specifies the azimuthal order of the modulation, \(\Omega=2\pi f_m\), and \(f_m\) denotes the temporal modulation frequency. As illustrated in Fig.~\ref{fig:1}(b), the permittivity profile rotates continuously about the cylinder axis while preserving its \(m\)-fold azimuthal periodicity.

For the Floquet analysis, we treat the lattice approximately as two-dimensional (infinitely extended in $z$ direction). Equation ~\eqref{eq1} is rewritten in terms of its temporal Fourier components as:
\begin{equation}
\varepsilon(\theta,t)=\bar{\varepsilon}+i\varepsilon''+\varepsilon_{+}(\theta)e^{-i\Omega t}+\varepsilon_{-}(\theta)e^{i\Omega t},
\label{eq2}
\end{equation}
with \(\varepsilon_{\pm}(\theta)=\bar{\varepsilon}\,\delta\varepsilon e^{\pm im\theta}/2\). The unmodulated eigenmodes are first obtained using the plane-wave expansion (PWE) method. In this calculation, each eigenmode of the unmodulated lattice is expanded as:
\begin{equation}
\psi_{\sigma}(\mathbf{r})=\sum_{\mathbf{G}} a_{\sigma,\mathbf{G}} e^{i(\mathbf{k}+\mathbf{G})\cdot \mathbf{r}},
\label{eq3}
\end{equation}
where \(\mathbf{G}\) is the reciprocal lattice vector, \(\mathbf{k}\) is the Bloch wave vector, and \(a_{\sigma,\mathbf{G}}\) is the plane-wave coefficient of mode \(\sigma\), in which \(\sigma\) is a general mode index. 

For the two degenerate unmodulated modes of interest, denoted by \(\alpha\) and \(\beta\), the time-periodic modulation introduces the coupling:
\begin{equation}
g_{\alpha\beta}^{(\pm)} \propto \int \psi_{\alpha}^{*}(\mathbf{r})\,\varepsilon_{\pm}[\theta(\mathbf{r})]\,\psi_{\beta}(\mathbf{r})\,d^{2}\mathbf{r},
\label{eq4}
\end{equation}
The resulting eigenmode field of the modulated lattice can be expanded in Floquet harmonics as: 
\begin{equation}
\Psi(\mathbf{r},t)=\sum_{n,\sigma} c_{n\sigma}\psi_{\sigma}(\mathbf{r})e^{-i(\omega+n\Omega)t},
\label{eq5}
\end{equation}
where \(\mathbf{r}=(x,y)\) denotes the position vector in the transverse plane, \(n\) labels the Floquet channel, and \(c_{n\sigma}\) is the expansion coefficient associated with the basis state indexed by \((n,\sigma)\). In this representation, the \(n\)-th Floquet channel corresponds to a frequency shifted by \(n\Omega\) relative to the frequency \(\omega\).

The eigenmodes are obtained by solving the full Floquet-Bloch eigenvalue problem including multiple Floquet channels \cite{park2021spatiotemporal}. 
For the $j$-th Floquet eigenmode, the population of the $n$-th Floquet channel is evaluated by summing the squared amplitudes of all unmodulated basis modes within that channel:
\begin{equation}
P_n^{(j)}
=
\sum_{\sigma}
\left|
c_{n\sigma}^{(j)}
\right|^2 .
\label{eq6}
\end{equation}

The normalized channel-resolved population is then defined as
\begin{equation}
\widetilde{P}_n^{(j)}
=
\frac{
P_n^{(j)}
}{
\sum_{n'=-N_F}^{N_F}
P_{n'}^{(j)}
}.
\label{eq7}
\end{equation}

Here, $N_F$ denotes the Floquet truncation order, set to $N_F=4$. 
The sums over $\sigma$ and $n$ run over the basis modes and Floquet channels included in the calculation, respectively. 
This population quantifies how the modal amplitude of a Floquet eigenstate is distributed among different frequency-shifted Floquet channels.

Using the above Floquet-Bloch framework, we analytically calculate the Floquet-channel populations of the Floquet eigenmodes. As shown in Fig.~\ref{fig:1}(c), the normalized zeroth-order Floquet-channel population \(\widetilde{P}_0\) at the \(\Gamma\) point identifies a doubly degenerate eigenfrequency in the static system. The right panel of Fig.~\ref{fig:1}(c) shows the corresponding band dispersion along the \(\Gamma\)-X direction, where the two branches are degenerate at \(\Gamma\) and split away from \(\Gamma\). When the spatiotemporal modulation is introduced (\(f_m = 100\ \mathrm{THz}\)), this degenerate doublet is lifted and forms two chiral branches, labeled by \(\ell=+1\) and \(\ell=-1\) in Fig.~\ref{fig:1}(d)--(e). Here, \(\ell=\pm1\) is used as a shorthand label for the two counter-rotating combinations of the original degenerate doublet; its interpretation as the orbital angular-momentum index will be formalized below. Figure~\ref{fig:1}(d) shows the \(\widetilde{P}_n\) of the two split branches for the \(n=-1,0,1\) orders, while Fig.~\ref{fig:1}(e) plots the corresponding frequency-resolved sideband spectra. The population is dominated by the zeroth-order component, with weaker first-order replicas generated by the modulation. Higher-order replicas can also arise in principle, but only the \(n=-1,0,1\) orders are displayed for clarity.

\begin{figure}[tb]
\centering\includegraphics[width=\linewidth]{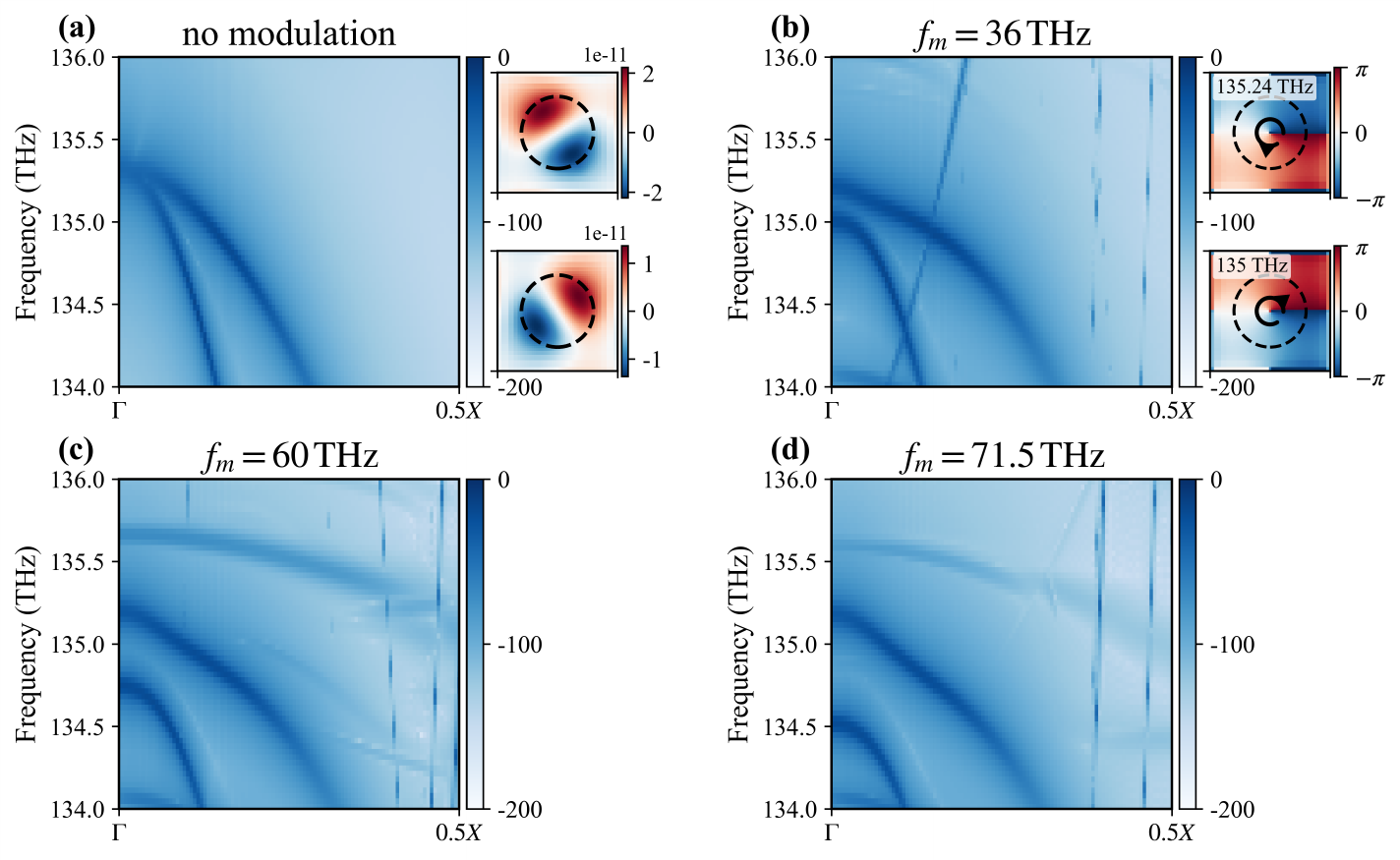}
\caption{
Band evolution of the spatiotemporal PTC along the \(\Gamma\)-X direction. 
(a) Band structure in the absence of modulation. Two nearly degenerate dipole-like modes appear at the \(\Gamma\) point around \(135.3\ \mathrm{THz}\). The inset shows the \(E_z\)-field distribution of the pair of degenerate dipole-like modes at \(135.3\ \mathrm{THz}\). 
(b)--(d) Band structures under modulation frequencies of \(36\ \mathrm{THz}\), \(60\ \mathrm{THz}\), and \(71.5\ \mathrm{THz}\), respectively. The modulation lifts the near degeneracy of the two dipole-like modes and opens a frequency gap at the \(\Gamma\) point. The gap increases with increasing modulation frequency. Insets in (b) show the phase distributions of the dipole-like modes associated with the upper and lower split bands, respectively, revealing their opposite handedness.The color scale represents the normalized Fourier amplitude of the monitored fields, defined as \(20\log_{10}(A/A_{\max})\), where \(A\) is the FFT amplitude at each \((k,f)\) point and \(A_{\max}\) is the maximum amplitude within the corresponding dataset.
}
\label{fig:2}
\end{figure}

\textit{Band structures and eigenmodes}---We carry out full-wave numerical simulations of the PTC using a finite-difference time-domain solver Tidy3D \cite{tidy3D}. Figure~\ref{fig:2}(a) presents the band structure of the unmodulated photonic crystal along the \(\Gamma\)-X direction. A pair of degenerate modes appears at the \(\Gamma\) point at \(135.3\ \mathrm{THz}\). The two insets display the eigen \(E_z\)-field distribution of the pair of degenerate modes, featuring two oppositely signed lobes characteristic of a dipole-like field distribution. The dashed circle marks the dielectric cylinder. Under the modulation with frequency \(f_m = 36\ \mathrm{THz}\), the degeneracy at the \(\Gamma\) point is lifted, as shown in Fig.~\ref{fig:2}(b). The two resulting branches are separated in frequency, with their midpoint shifted from the original degeneracy frequency of \(135.3\ \mathrm{THz}\) due to the modulation-induced modification of the effective optical response of the PTC. The \(E_z\)-phase distributions in the insets reveal that the mode at \(135.24\ \mathrm{THz}\) rotates counterclockwisely, whereas that at \(135\ \mathrm{THz}\) rotates clockwisely. Figure~\ref{fig:2}(c) and 2(d) show the bands and the eigen \(E_z\) fields at the modulation frequencies of \(f_m = 60\ \mathrm{THz}\) and \(f_m = 71.5\ \mathrm{THz}\), respectively. Increasing \(f_m\) further enlarges the splitting between the two formerly degenerate branches. At \(f_m = 60\ \mathrm{THz}\), the modes are located at \(135.24\) and \(134.75\ \mathrm{THz}\), while at \(f_m = 71.5\ \mathrm{THz}\), they occur at \(135.24\) and \(134.51\ \mathrm{THz}\). Despite the enlarged splitting, the higher-frequency mode consistently exhibits counterclockwise rotation, whereas the lower-frequency mode rotates in clockwise direction. The full band structures are shown in Fig. S1 of Supplementary Materials.

\begin{figure}[tb]
\centering\includegraphics[width=\linewidth]{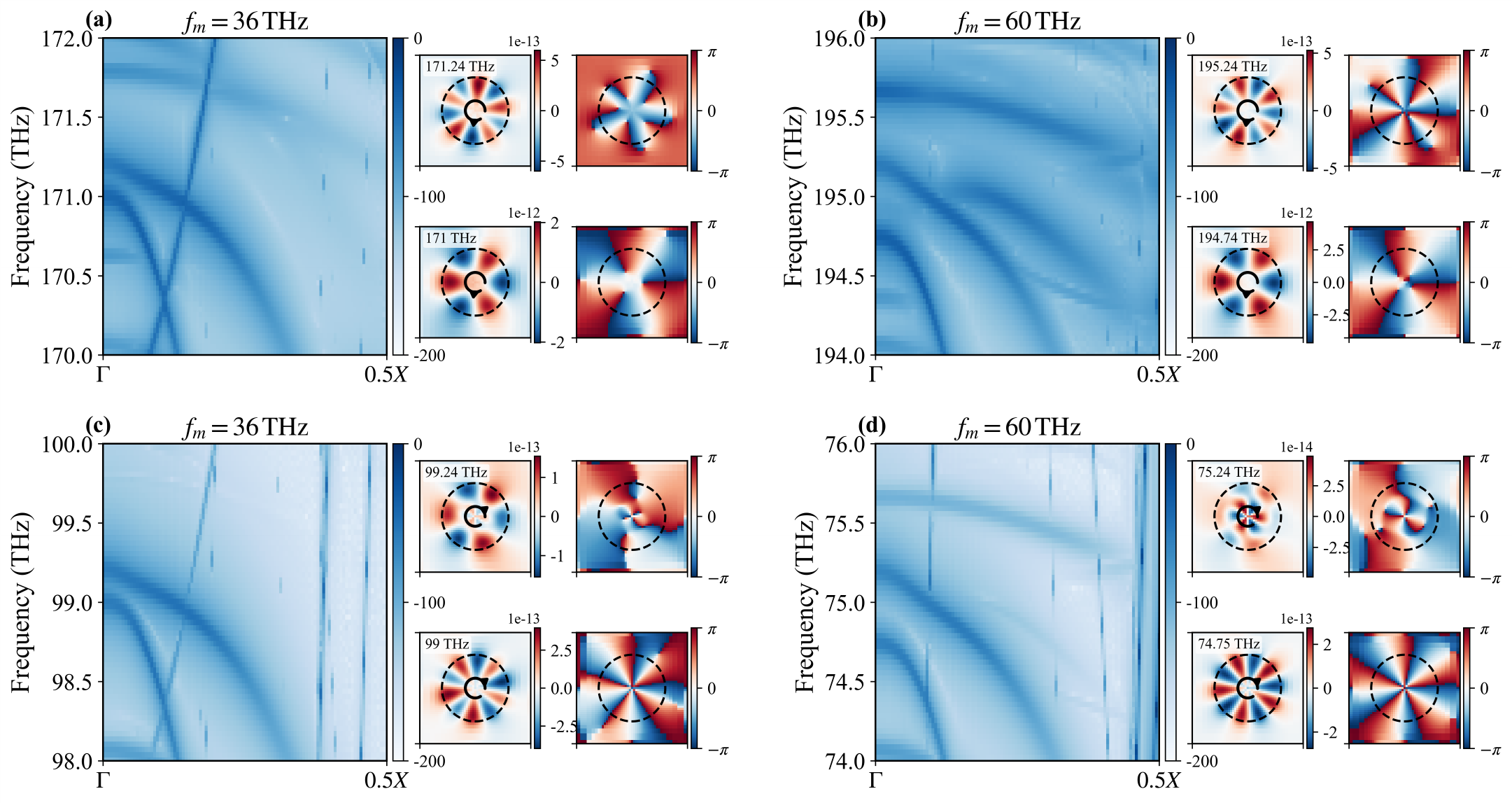}
\caption{
Floquet-sideband band structures and mode profiles. Floquet-sideband band structures at modulation frequencies of \(36\ \mathrm{THz}\) (a),(c) and \(60\ \mathrm{THz}\) (b),(d). Panels (a) and (b) show the \(n=+1\) Floquet sideband replicated from the dipole-like mode, whereas panels (c) and (d) show the corresponding \(n=-1\) Floquet sideband. The four subpanels in each panel display the field characteristics of the split bands at the \(\Gamma\) point. The two left subpanels show the \(E_z\) fields of the higher-order modes evolved from the dipole-like modes under modulation; the two right subpanels show the corresponding phase distributions. The dashed circles mark the characteristic cylinder region.
}

\label{fig:3}
\end{figure}

Figure~\ref{fig:3} shows the rotational characteristics of the first-order (\(n=\pm1\)) Floquet sidebands. Figures~\ref{fig:3}(a) and \ref{fig:3}(b) present the \(n=+1\) sidebands for \(f_m=36\ \mathrm{THz}\) and \(60\ \mathrm{THz}\), respectively, whose frequencies are shifted upward by \(f_m\) relative to the corresponding zeroth-order modes. Figures~\ref{fig:3}(c) and \ref{fig:3}(d) show the associated \(n=-1\) sidebands, shifted downward by \(f_m\). In each panel, the upper and lower insets correspond to the higher- and lower-frequency modes, respectively. The left subpanels display the \(E_z\)-field distributions, whereas the right subpanels show the corresponding phase maps. The field and phase distributions reveal that both modes in the \(n=+1\) sideband rotate counterclockwisely, whereas both modes in the \(n=-1\) sideband rotate clockwisely. Thus, the two modes within each sideband rotate in the same direction, whereas the \(n=+1\) and \(n=-1\) sidebands rotate in opposite directions. This behavior originates from the simultaneous frequency and orbital angular-momentum conversion induced by the modulation. To formalize the chiral labels \(\ell=\pm1\) used above, the spatiotemporal dependence of a mode can be written as:
\begin{equation}
\psi_{\ell}(\theta,t)\propto e^{i(\ell\theta-\omega t)},
\label{eq8}
\end{equation}
where \(\theta\) increases in the counterclockwise direction and \(\ell\) denotes the orbital angular-momentum quantum number. Under this convention, positive and negative \(\ell\) correspond to counterclockwise and clockwise azimuthal phase winding, respectively, and hence to the temporal rotation of the field patterns. In the unmodulated structure, the two degenerate dipole-like modes can be combined to form counter-rotating states labeled by \(\ell=+1\) and \(\ell=-1\). Upon introducing the counterclockwise fourfold modulation, this degeneracy is lifted for the zeroth-order Floquet modes. As established in Fig.~\ref{fig:2}, the higher- and lower-frequency zeroth-order modes exhibit counterclockwise and clockwise rotations, respectively, and are therefore assigned \(\ell_{0}=+1\) and \(\ell_{0}=-1\).

The rotating permittivity perturbation can be decomposed as:
\begin{equation}
\delta\varepsilon(\theta,t)\propto \cos(m\theta-\Omega t)
=\frac{1}{2}\left[e^{i(m\theta-\Omega t)}+e^{-i(m\theta-\Omega t)}\right].
\label{eq:perturbation_decomposition}
\end{equation}
The two spatiotemporal harmonics mediate simultaneous frequency and angular-momentum conversion characterized by \((\delta \ell,\delta \omega)=(\pm m,\pm \Omega)\), respectively. Consequently, the first-order Floquet sidebands satisfy
\begin{equation}
\omega_{\pm1}=\omega_0\pm\Omega,\qquad \ell_{\pm1}=\ell_0\pm m,
\label{eq:sideband_relation}
\end{equation}
where \(\omega_0\) and \(\ell_0\) are the frequency and orbital angular-momentum index of the corresponding zeroth-order mode. The modulation in Fig.~\ref{fig:1} has \(m=4\). For the \(n=+1\) sideband, the states with \(\ell_0=+1\) and \(\ell_0=-1\) are converted into states with \(\ell_{+1}=+5\) and \(\ell_{+1}=+3\), respectively. Both indices are positive, accounting for the counterclockwise rotation of both \(n=+1\) sideband modes. For the \(n=-1\) sideband, the states have \(\ell_{-1}=-3\) and \(\ell_{-1}=-5\). Their negative signs account for the clockwise rotation of both \(n=-1\) sideband modes, consistent with the field and phase distributions in Fig.~\ref{fig:3}.

\begin{figure}[tb]
\centering\includegraphics[width=\linewidth]{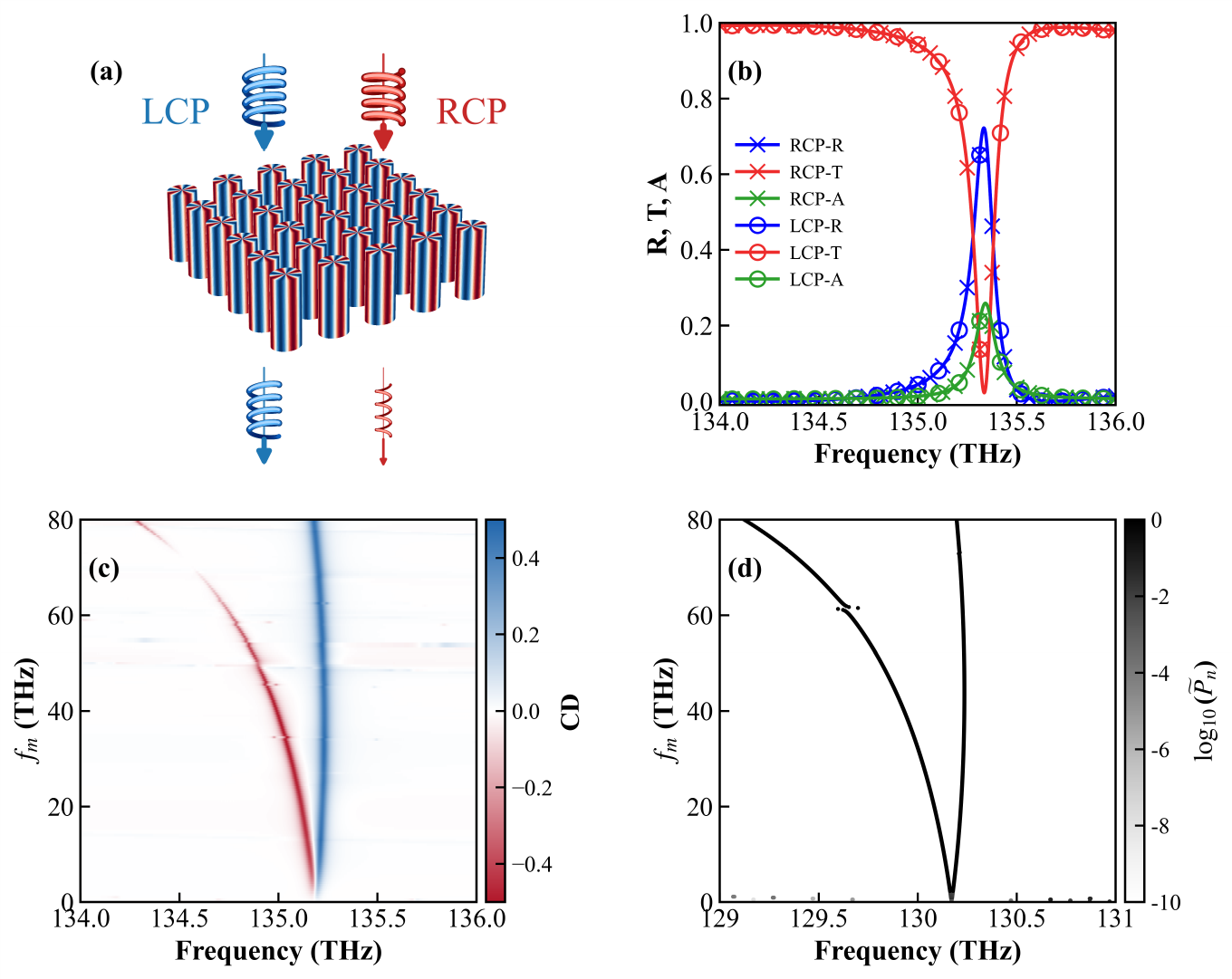}
\caption{
Circularly polarized optical response. (a) Schematic illustration of the optical response of the PTC to LCP and RCP light. (b) Reflectance, transmittance and absorptance spectra of LCP and RCP in the absence of modulation. (c) Frequency-resolved CD as a function of \(f_m\), revealing the emergence and evolution of chiral optical response under modulation. (d)  \(\widetilde{P}_n\) at the \(\Gamma\) point as a function of \(f_m\), analytically calculated using the simplified two-dimensional model described in the \textit{Theory} section.
}

\label{fig:4}
\end{figure}

\textit{Chiral optical response}---The intrinsic chirality of the orbital angular momentum modes can give rise to chiral optical response. To understand this, we simulate the reflectance \(R\), transmittance \(T\), and absorbance \(A\) of the PTC under normally incident circularly polarized plane wave illumination at different \(f_m\), as schematically shown in Fig.~\ref{fig:4}(a). The corresponding spectra are provided in Fig.~S2 of the Supplementary Material. In the absence of modulation (\(f_m = 0\)), the spectra for LCP and RCP incidence are identical, as shown in Fig.~4(b). In particular, the resonant features of the two polarizations coincide at 135.3~THz, indicating that the static structure preserves the degeneracy between the two circularly polarized channels, consistent with the degenerate bands at the \(\Gamma\) point shown in Fig.~\ref{fig:2}(a). The spectra show two branches due to the resonant modes carrying opposite angular momentum, which exhibit distinct chiral responses under the excitation of circularly polarized light. To quantify the resulting chiral response, we evaluate the circular dichroism defined as \(CD = A_{\mathrm{LCP}} - A_{\mathrm{RCP}}\), which develops into two branches of opposite sign as \(f_m\) increases. The lower-frequency branch exhibits negative CD, whereas the higher-frequency branch exhibits positive CD, consistent with the eigenmode properties in Fig.~\ref{fig:2}. The splitting is asymmetric, with the RCP-associated branch shifting further towards lower frequency than the LCP-associated branch shifts towards higher frequency over the same modulation range. Meanwhile, the spectral separation between the two branches increases progressively with \(f_m\), establishing modulation as an efficient route to continuously tune the chiral optical response. To clarify the origin of the observed branch splitting, we analytically calculate the \(\widetilde{P}_n\) at the \(\Gamma\) point using the simplified two-dimensional model described in the \textit{Theory} section. As shown in Fig.~\ref{fig:4}(d), the originally degenerate modes split into two branches as \(f_m\) increases. The asymmetric splitting and the resulting branch trajectories agree well with the CD features obtained from the full-wave simulations in Fig.~\ref{fig:4}(c), confirming that the chiral optical response originates from modulation-induced splitting of the degenerate angular-momentum modes. The slight deviations from the smooth branch trajectories observed at certain modulation frequencies can be attributed to weak hybridization with nearby Floquet replicas from other orders, which locally perturbs the split \(n = 0\) modes. It is also observed in Fig.~\ref{fig:4}(c) that, near \(f_m = 71.5~\mathrm{THz}\), the RCP-coupled branch becomes strongly suppressed and eventually disappears. This behavior indicates that, under this modulation condition, the coupling between the incident field and the corresponding radiative resonant channel is significantly inhibited. The suppression can be qualitatively understood as arising from a modulation-induced reconfiguration of the multipole modes, which reduces the net far-field radiation through destructive interference.

\textit{Conclusion}---In summary, we demonstrate that a PTC can emulate rotational dynamics in a geometrically static structure, lifting a static chiral modal degeneracy and producing nondegenerate counter-rotating eigenstates carrying opposite orbital angular momenta. This synthetic rotation enables selective coupling to LCP and RCP light, giving rise to tunable chiral response such as CD that can be actively controlled by the modulation frequency. The spatiotemporal modulation induces orbital angular-momentum conversion among Floquet replica bands, providing a mechanism for achieving dynamic chiral response without relying on mechanical motion. Our results establish spatiotemporal modulation as a versatile route for engineering chiral light–matter interactions and suggest that analogous concepts may extend to other wave platforms. More broadly, this work shows that spatiotemporal symmetry engineering can complement and even surpass spatial structure design, opening avenues for reconfigurable chiral and nonreciprocal photonic devices.

\textit{Acknowledgments}---The work described in this paper was supported by grants from the National Natural Science Foundation of China (No. 12322416) and the Research Grants Council of the Hong Kong Special Administrative Region, China (Project No. AoE/P-502/20).




\bibliography{sample}

\end{document}